\documentclass[english,aps,prl,reprint,superscriptaddress]{revtex4-1}
\usepackage[T1]{fontenc}
\usepackage[latin9]{inputenc}
\usepackage{geometry}
\usepackage{amsmath}
\usepackage{amssymb}
\usepackage{graphicx}

\allowdisplaybreaks

\providecommand{\tabularnewline}{\\}

\makeatother

\usepackage{babel}

\newcommand*{\citen}[1]{%
  \begingroup
    \romannumeral-`\x 
    \setcitestyle{numbers}%
    \cite{#1}%
  \endgroup   
}

\makeatletter

\begin{document}
\title{Polariton-assisted Singlet Fission in Acene Aggregates}
\author{Luis A. Mart\'inez-Mart\'inez}
\affiliation{Department of Chemistry and Biochemistry, University of California
San Diego, La Jolla, California 92093, United States}
\author{Matthew Du}
\affiliation{Department of Chemistry and Biochemistry, University of California
San Diego, La Jolla, California 92093, United States}
\author{Raphael F. Ribeiro}
\affiliation{Department of Chemistry and Biochemistry, University of California
San Diego, La Jolla, California 92093, United States}
\author{St\'ephane K\'ena-Cohen}
\affiliation{Department of Engineering Physics, \'Ecole Polytechnique de Montr\'eal,
Montr\'eal H3C 3A7, QC, Canada}
\author{Joel Yuen-Zhou}
\affiliation{Department of Chemistry and Biochemistry, University of California
San Diego, La Jolla, California 92093, United States}





\begin{abstract}
Singlet fission is an important candidate to increase energy conversion
efficiency in organic photovoltaics by providing a pathway to increase
the quantum yield of excitons per photon absorbed in select materials.
We investigate the dependence of exciton quantum yield for acenes
in the strong light-matter interaction (polariton) regime, where the
materials are embedded in optical microcavities. Starting from an
open-quantum-systems approach, we build a kinetic model for time-evolution
of species of interest in the presence of quenchers and show that
polaritons can decrease or increase exciton quantum yields compared
to the cavity-free case. In particular, we find that hexacene, a typically
poor singlet-fission candidate, can feature a higher yield than cavity-free
pentacene when assisted by polaritonic effects. Similarly, we show
that pentacene yield can be increased when assisted by polariton states.
Finally, we address how various relaxation processes between bright
and dark states in lossy microcavities affect polariton photochemistry.
Our results also provide insights on how to choose microcavities to
enhance similarly related chemical processes.
\end{abstract}

\pacs{Strong light-matter coupling, pentacene, tetracene, dark states,
microcavity, dynamics }

\maketitle
\setlength{\abovedisplayskip}{3pt} 
\setlength{\belowdisplayskip}{3pt}

Singlet fission (SF) is a spin-allowed process undergone by select
materials that permits the conversion of a singlet exciton into a
triplet-triplet (TT) state with an overall singlet character, which
later decoheres and forms two triplet excitons. This process has been
used to enhance the external quantum efficiency of organic solar cells
\cite{Congreve2013,Yost2014} by allowing a single absorbed photon
to produce more than one exciton. In this work we explore the influence
of strong light-matter coupling (SC) on the TT yield of acenes. This
regime can be achieved at room-temperature, for example, in optical
microcavities enclosing densely packed organic dyes \cite{Hutchison2012}.
Under these conditions, the energy of interaction between the microcavity
photonic modes and the molecular degrees of freedom of the material
is larger than their respective linewidths. The hybrid states that
arise from this interaction are called \emph{polaritons}. The latter
have previously been exploited to tune the properties and functionality
of organic materials at the molecular level. For instance, there have
been experimental and theoretical efforts to explore the potential
applications of SC in photochemistry, where the electrodynamic vacuum
can play a role in molecular processes \cite{Hutchison2012,Herrera2016,Galego2016,Flick2016,Thomas2016,Zhong2016,Ebbesen2016,Zhong2017,Martinez2017}.
Previous studies have explored SC in the context of exciton harvesting
and transport \cite{Gonzalez-Ballestero2015,Feist2015}, Raman scattering\cite{Strashko2016,delPino2015}
and photoluminescence spectroscopy \cite{Herrera2016a,Melnikau2016},
Bose-Einstein condensation \cite{Andre2006,Gerace2012,Nguyen2015},
and topologically-protected states \cite{Yuen2016}, just to mention
a few examples.

By developing a microscopic model for the relevant processes, we address
the effects of SC on the TT yield in aggregates of acene dyes (tetracene,
pentacene and hexacene) and determine the important molecular parameters
that rule this yield. Our starting point is a kinetic model based
on a Pauli master equation formalism that describes the population
dynamics of the states that take part in SF \cite{nitzan2006}. We
then use this model to elucidate the circumstances under which polaritons
can enhance SF under realistic dissipative conditions.

\emph{Theoretical model}.\textemdash{} We consider a simplified one-dimensional
acene aggregate comprised of $N$ identical molecules embedded in
a microcavity and strongly interacting with a single electromagnetic
mode supported by the latter. The Hamiltonian of the model is given
by 
\begin{multline}
H  =H_{S}+H_{B}+H_{S-B}+H_{p}+H_{p-S}+H_{TT}+H_{TT-B}+ H_{TT-S},\label{gral_Ham}
\end{multline}
where $H_{S}$ ($H_{TT}$) is the electronic singlet (TT) Hamiltonian
of the aggregate given by ($\hbar=1$)

\begin{subequations}

\begin{align}
H_{S} & =\sum_{n}\overline{\omega}_{e}|n\rangle\langle n|,\label{eq:singlet}\\
H_{TT} & =\sum_{n=0}^{N-1}\overline{\omega}_{TT}|T_{n}T_{n+1}\rangle\langle T_{n}T_{n+1}|,\label{eq:TT}
\end{align}

\end{subequations}\raggedbottom
\noindent where $|n\rangle$ is a localized singlet
(Frenkel) exciton \cite{Frenkel1931} at the $n$th site (molecule),
and $|T_{n}T_{n+1}\rangle$ denotes a TT state delocalized over sites
$n$ and $n+1$. Here $\overline{\omega}_{e}=\omega_{e}+\sum_{i}\omega_{i}\lambda_{S,i}^{2}$
($\overline{\omega}_{TT}=\omega_{TT}+2\sum_{i}\omega_{i}\lambda_{T,i}^{2}$)
is the vertical singlet (TT) excitation frequency, where $\omega_{e}$
($\omega_{TT}$) and $\lambda_{S,i}$ ($\lambda_{T,i}$) are the 0-0
excitation frequency and the square root of the Huang-Rhys factor
\cite{Volkhard2008} for the $i$th vibrational mode coupled to the
transition $|G\rangle\rightarrow|n\rangle$ ($|G\rangle\rightarrow|T_{n}T_{n+1}\rangle$),
respectively, and $|G\rangle$ is the state corresponding to all molecules
in the electronic ground state. $H_{B}=\sum_{n,i}\omega_{i}b_{n,i}^{\dagger}b_{n,i}$
accounts for the vibrational degrees of freedom of the ensemble, where
$b_{n,i}^{\dagger}$ ($b_{n,i}$) is the creation (annihilation) operator
of the $i$-th harmonic vibrational degree of freedom with frequency
$\omega_{i}$ on site $n$. The singlet (TT) vibronic couplings are
encoded in $H_{S-B}$ ($H_{TT-B}$), given by

\begin{subequations}

\begin{align}
H_{S-B} & =\sum_{n,i}|n\rangle\langle n|\omega_{i}\lambda_{S,i}(b_{n,i}+\text{h.c.}),\label{eq:HSB}\\
H_{TT-B} & =\sum_{n=0}^{N-1}|T_{n}T_{n+1}\rangle\langle T_{n}T_{n+1}|\nonumber \\
\times & \sum_{i}\omega_{i}\lambda_{T,i}(b_{n,i}+b_{n+1,i}+\text{h.c.}).\label{eq:HTTB}
\end{align}

\end{subequations}

The singlet-TT electronic coupling is (assuming periodic boundary
conditions $|T_{-1}T_{0}\rangle=|T_{N-1}T_{0}\rangle$): \cite{Teichen2015}
\begin{equation}
H_{TT-S}=\frac{V_{TT-S}}{2}\sum_{n=0}^{N-1}\Big[\left(|T_{n}T_{n+1}\rangle+|T_{n-1}T_{n}\rangle\right)\langle n|+\text{h.c.}\Big].\label{eq:HTTS}
\end{equation}
Finally, the photonic degree of freedom is included in $H_{p}=\omega_{ph}a^{\dagger}a$
where $a^{\dagger}$($a$) is the creation (annihilation) operator
of the cavity photonic mode. Its interaction with the singlet excitons
is described by the light-matter Hamiltonian
\begin{align}
H_{p-S} & =\sum_{n}g\big(a^{\dagger}|G\rangle\langle n|+\text{h.c.}\big)\nonumber \\
 & =\sqrt{N}g\Big(a^{\dagger}|G\rangle\langle k=0|+\text{h.c.}\Big)\label{eq:collect_coup}\\
 & =\frac{\Omega}{2}\Big(a^{\dagger}|G\rangle\langle k=0|+\text{h.c.}\Big)\nonumber 
\end{align}
where in the second line we have introduced a delocalized Fourier
basis for the singlet excitons $|k\rangle=\frac{1}{\sqrt{N}}\sum_{n}e^{ikn}|n\rangle$
$k=\frac{2\pi m}{N},\,m=0,1,2,\dots,N-1$. The $\sqrt{N}g$ term in
(\ref{eq:collect_coup}) is the collective light-matter coupling and
$\Omega$ is the so-called Rabi splitting. Importantly, the singlet
excitons are optically bright, in contrast to the dark TT states \cite{Teichen2015}.
Because of this property, our model does not feature a TT term analogous
to Eq. (\ref{eq:collect_coup}). 

\begin{figure*}
\includegraphics[width=0.8\textwidth]{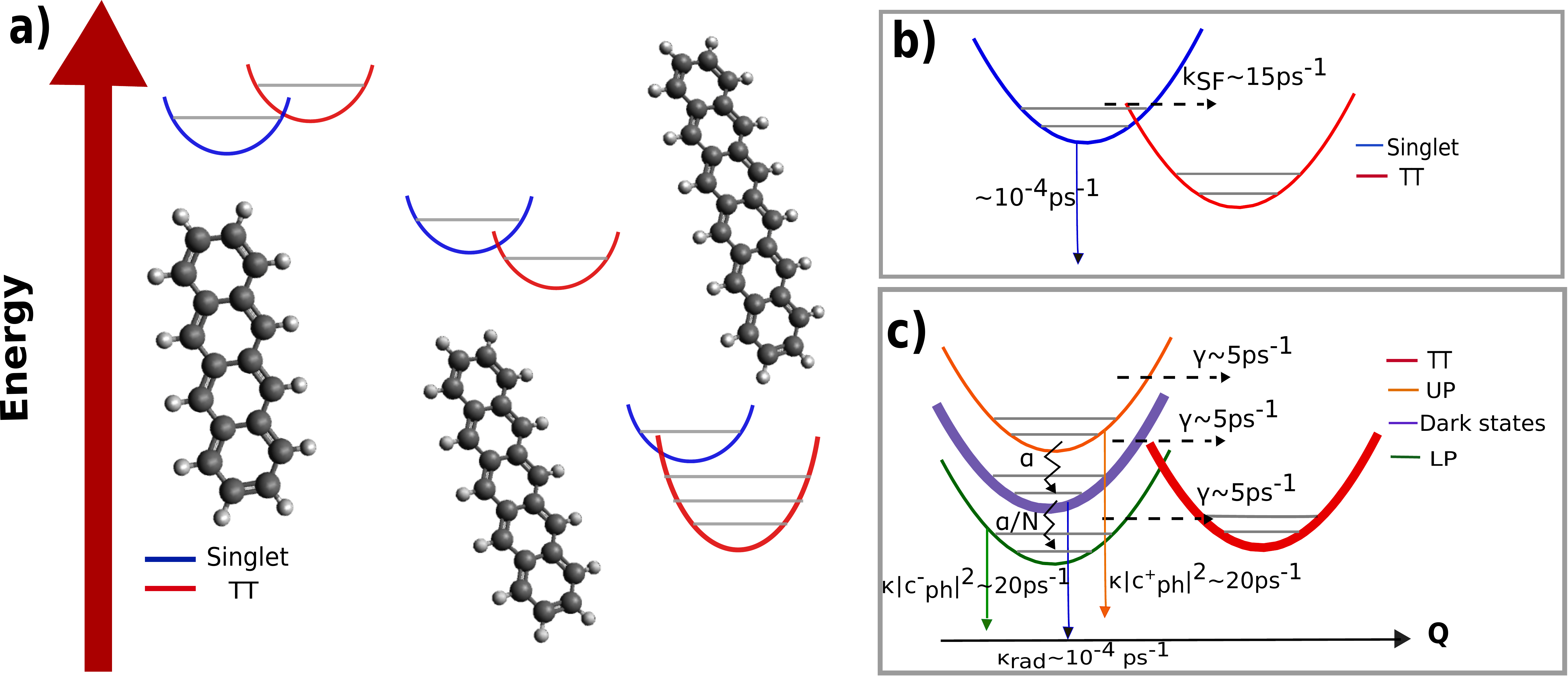}

\caption{a) Bare triplet-triplet (red) and singlet (blue) vibronic energies
of the different molecules considered. From left to right: tetracene,
pentacene and hexacene. For clarity only the vibrational mode with
highest frequency is shown. The SF dynamics is schematically shown
for bare case (b) and the SC scenario (c). In (c) the fastest decay
constant $\alpha\approx100$ ps$^{-1}$ is due to vibrational relaxation,
$\gamma$ is the dressed SF rate, $\kappa$ is the cavity-photon leakage
rate and $|c_{ex}^{+}|^{2}$($|c_{ex}^{-}|^{2}$) is the exciton fraction
in the upper (lower) polariton. $k_{f}$ is the singlet fluorescence
decay rate. Continuous arrows denote radiative decay. Thicker lines
indicate larger density of states. Approximate SF and relaxation rates
for pentacene interacting with a resonant photonic mode are included.
The decay timescale for the triplet-triplet state is significantly
longer than the timescales considered here.\label{fig:decay_scheme}}
\end{figure*}

\vspace{-2pt} In our approach we consider the reduced population
dynamics of the manifold of TT states $\left(\{|T_{n}T_{n+1}\rangle\}\right)$,
the two polariton states ($|\pm\rangle$) and the so-called \emph{dark
states} ($\{|d\rangle=|k\neq0\rangle\}$) that emerge from SC (see
Fig. \ref{fig:decay_scheme} and the Supporting Information (SI) for
additional details of the method), where

\begin{subequations}\label{polaritons}

\begin{align}
|+\rangle & =c_{ph}^{+}|G\rangle\otimes|1_{ph}\rangle+c_{ex}^{+}|k=0\rangle\otimes|0_{ph}\rangle,\label{eq:UP}\\
|-\rangle & =c_{ph}^{-}|G\rangle\otimes|1_{ph}\rangle+c_{ex}^{-}|k=0\rangle\otimes|0_{ph}\rangle.\label{eq:LP}
\end{align}

\end{subequations}\noindent with (zeroth-order) eigenenergies given
by
\begin{equation}
\omega_{\pm}=\frac{\overline{\omega}_{e}+\omega_{ph}}{2}\pm\sqrt{\bigg(\frac{\overline{\omega}_{e}-\omega_{ph}}{2}\bigg)^{2}+\big(\sqrt{N}g\big)^{2}},\label{eq:pol_eigenenergies}
\end{equation}
In Eq. (\ref{polaritons}), $|n_{ph}\rangle$ is the state with $n$
photons in the photonic space and $|+\rangle$ ($|-\rangle$) is the
upper (lower) polariton state. Meanwhile, $c_{ph}^{\pm}$ and $c_{ex}^{\pm}$
are the Hopfield coefficients for the photon and exciton components, respectively, 
of the polariton states \cite{Hopfield1958}. The reduced dynamics
is described by means of a Pauli master equation derived using the
Redfield formalism under the secular and Markov approximations \cite{nitzan2006}
(see SI). The kinetic model can be summarized by the following equations, 

\begin{subequations}\label{kin_pol}

\begin{align}
\partial_{t}P_{\pm}(t) & =-|c_{ex}^{\pm}|^{2}\frac{\alpha(\omega_{\pm D})}{N}(N-1)P_{\pm}+|c_{ex}^{\pm}|^{2}\frac{\alpha(\omega_{D\pm})}{N}P_{D}^{total}\label{kin_up}\\
 & -|c_{ex}^{\pm}|^{2}\frac{\gamma(\omega_{TT,\pm})}{N}(N)P_{\pm}+|c_{ex}^{\pm}|^{2}\frac{\gamma(\omega_{\pm,TT})}{N}P_{TT}^{total}\nonumber \\
 & -\left(|c_{ex}^{\pm}|^{2}k_{c}(\omega_{\pm})+|c_{ph}^{\pm}|^{2}k_{phot}\right)P_{\pm},\nonumber 
\end{align}
\begin{align}
\partial_{t}P_{D}^{total}(t)&=|c_{ex}^{+}|^{2}\frac{\alpha(\omega_{+D})}{N}\left(N-1\right)P_{+}\nonumber\\&-|c_{ex}^{+}|^{2}\frac{\alpha(\omega_{D+})}{N}P_{D}^{total}+|c_{ex}^{-}|^{2}\frac{\alpha(\omega_{-D})}{N}\left(N-1\right)P_{-}\nonumber\\&-|c_{ex}^{-}|^{2}\frac{\alpha(\omega_{D-})}{N}P_{D}^{total}-\gamma(\omega_{TT,D})P_{D}^{total}\label{kin_dark}\\&+\gamma(\omega_{D,TT})P_{TT}^{total}-k_{c}(\omega_{D})P_{D}^{total},\nonumber
\end{align}
\begin{align}
\partial_{t}P_{TT}^{total}(t) & =|c_{ex}^{+}|^{2}\gamma(\omega_{TT,+})P_{+}-|c_{ex}^{+}|^{2}\frac{\gamma(\omega_{+,TT})}{N}P_{TT}^{total}\label{kin_triplet}\\
 & -\gamma(\omega_{D,TT})P_{TT}^{total}+\gamma(\omega_{TT,D})P_{D}^{total}\nonumber \\
 & +|c_{ex}^{-}|^{2}\gamma(\omega_{TT,-})P_{-}-|c_{ex}^{-}|^{2}\frac{\gamma(\omega_{-,TT})}{N}P_{TT}^{total},\nonumber 
\end{align}

\end{subequations}
\raggedbottom
\noindent where $P_{+}(t)$ ($P_{-}(t)$) is the population in the $|+\rangle$
($|-\rangle$) state and $P_{D}^{total}$ ($P_{TT}^{total}(t)$) is
the total population in the dark (TT) state manifold. The photon and
exciton content of the polaritonic states are given by $|c_{ph}^{\pm}|^{2}$
and $|c_{ex}^{\pm}|^{2}$, while $\omega_{D}=\overline{\omega}_{e}$
and $\omega_{ab}=\omega_{a}-\omega_{b}$. In Eqs. (\ref{kin_up}) and (\ref{kin_dark}),
we phenomenologically introduce the rate constant $k_{c}(\omega)$
to account for the contribution to the decay rate of the dressed states
due to their singlet exciton fraction. $k_{c}(\omega)$ can account
for\emph{ e.g.} radiative or non-radiative relaxation of the singlet
to the ground electronic state , or its conversion into charges at
the interface with charge acceptors, which is the case in donor-acceptor
blends used for organic solar cells. The various cases are treated
in more detail below.

The $\alpha(\omega)$ transfer rates appearing in Eqs. (\ref{kin_up}) and (\ref{kin_dark})
are calculated in terms of a bath spectral density and thermal populations
at frequency $\omega$. The $\gamma(\omega)$ rates are computed with
a Bixon-Jortner-like \cite{Jortner_Bixon1988} equation adapted to
the SC regime. For sake of simplicity, in the calculation of the latter
the vibrational bath is treated by using an effective high (low) frequency
$\overline{\omega}_{h}$ ($\overline{\omega}_{l}$) that satisfies
$\overline{\omega}_{h}\gg1/\beta$ ($\overline{\omega}_{l}\ll1/\beta$),
to which we associate a so-called inner (outer) sphere reorganization
energy \cite{jortner1976}. We refer the reader to the SI for details
of the derivation and other relevant parameters employed in the calculation
of $\alpha(\omega)$ and $\gamma(\omega)$. The parameters $\Delta G=\omega_{TT}-\omega_{e}$,
and $V_{TT-S}/2$ take part in these calculations and are treated
as material-dependent; they are taken from Ref. \citen{Yost2014} and
are summarized in Table \ref{tab:param}. 

\begin{figure}
\includegraphics[scale=0.65]{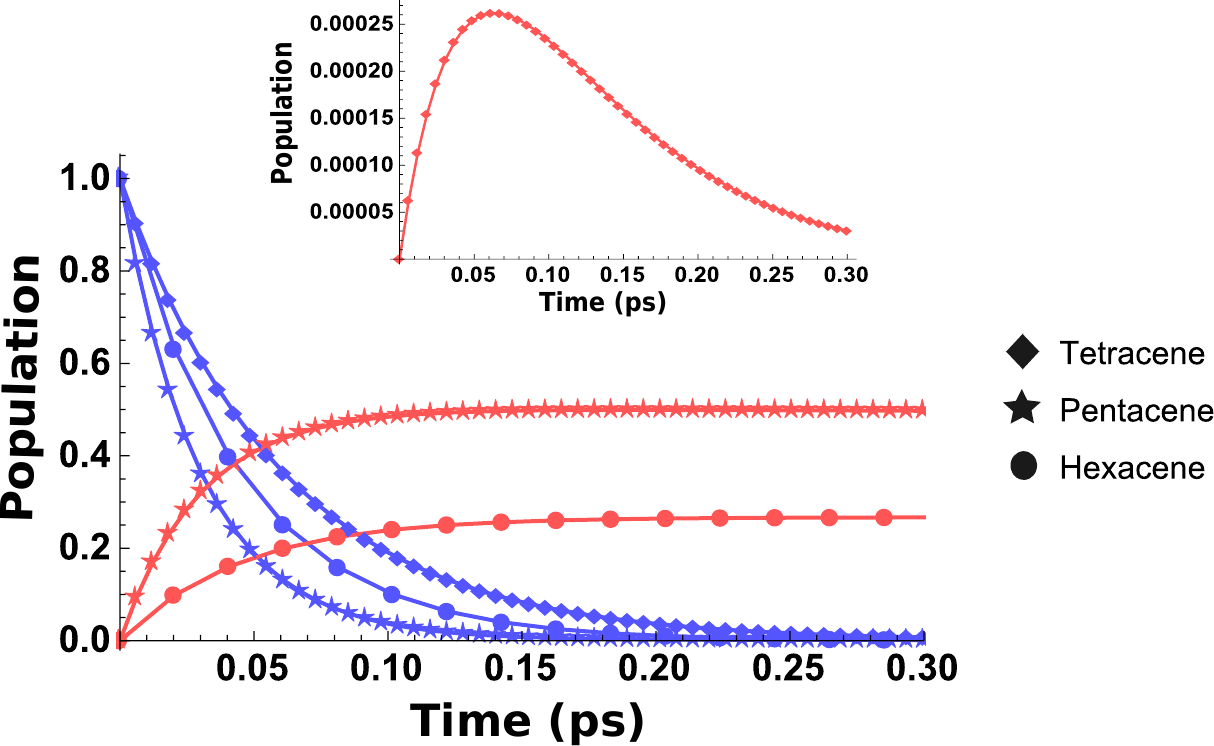}

\caption{Time evolution of populations in the singlet (blue) and TT electronic
states (red) of the molecules considered in this work in the bare
case. We consider the initial conditions that follow from pumping
the singlet state at $t=0$ ($P_{S}(t=0)=1$, $P_{TT}(t=0)=0$), for
all the molecules in question. Inset: time evolution of $P_{TT}(t)$
for tetracene. \label{fig:bare_dyn}}
\end{figure}

\begin{table}
\begin{tabular}{|c|c|c|c|c|}
\hline 
Molecule & $V_{TT-S}/2$ (meV) & $\Delta G$ (meV) & $k_{SF}$ ($ps^{-1}$) & $\epsilon_{TT}$ ($\%$)\tabularnewline
\hline 
\hline 
Tetracene & 41.5 & 150 & 0.01 & 0.01\tabularnewline
\hline 
Pentacene & 42 & -110 & 17 & 100\tabularnewline
\hline 
Hexacene & 22 & -630 & 6 & 48\tabularnewline
\hline 
\end{tabular}

\caption{Summary of bare material-dependent parameters. The values for $\frac{V_{TT-S}}{2}$
and $\Delta G=\omega_{TT}-\omega_{e}$ were taken from Ref. \cite{Yost2014}.
Based on these, we calculate the bare SF rates $k_{SF}$ and bare
TT yields $\epsilon_{TT}$. Importantly, $\epsilon_{TT}$ is calculated
assuming that SF competes with the fast singlet decay process (charge
production) $k_{CT}=17$ ps$^{-1}$, as discussed in the main text.\label{tab:param}}
\end{table}
\emph{Discussion of results}.\textemdash{} We first consider the population
dynamics of bare (cavity-free) systems containing one of the acene
molecules,

\begin{subequations}

\begin{align}
\frac{dP_{S}(t)}{dt} & =-\big(k_{SF}+k_{c}(\omega_{D}))P_{S}(t)+k_{TF}P_{TT}(t)\label{eq:PS}\\
\frac{dP_{TT}(t)}{dt} & =k_{SF}P_{S}(t)-k_{TF}P_{TT}(t)\label{eq:PTT}
\end{align}

\end{subequations}\noindent where $P_{S}(t)$ ($P_{TT}(t)$) is the
population of the singlet (TT) electronic state of a given acene,
$k_{SF}$ is the bare SF rate, and $k_{TF}$ is the bare triplet fusion
rate, which corresponds to the reverse process to SF. $k_{SF}$ and
$k_{TF}$ are calculated by means of the Bixon-Jortner equation \cite{Jortner_Bixon1988},
with the parameters in Table (\ref{tab:param}). For all cases we
compute $P_{S}(t)$ and $P_{TT}(t)$ assuming $P_{S}(0)=1$, $P_{TT}(0)=0$
(see Fig. \ref{fig:bare_dyn}). Under these conditions, we define
the TT yield

\begin{equation}
\epsilon_{TT}(t^{*}\gg0)=200\%\times P_{TT}(t^{*})\label{eq:TT_yield}
\end{equation}
as the relevant figure of merit for our subsequent analysis where
$t^{*}$ was chosen to reach a stationary $P_{TT}(t)$ value for pentacene
and hexacene. We notice that when $k_{c}(\omega)=k_{f}=O(10^{-4})$
ps$^{-1}$ (singlet fluorescence rate \cite{nitzan2006}) pentacene
and hexacene are expected to exhibit a $200\%$ TT yield in view of
$k_{SF}\gg k_{f}$ and $k_{TF}=e^{\beta\Delta G}k_{SF}\ll k_{SF}$
(detailed balance, where $\beta$ is the inverse temperature). This
contrasts with tetracene, in view of its higher TT energy compared
to the singlet, $\left(\omega_{TT}>\omega_{e}\right)$ so that $k_{SF}\ll k_{TF}$
and the TT population decays to zero for long $t$. The experimental
TT yield of tetracene is well above zero \cite{Burdett2010,Zhu2016},
which is in contrast with our findings (see Fig. \ref{fig:bare_dyn}).
The reason is that the mechanism that leads to this unexpected yield
is related to entropic gain not considered in our model. \cite{Chan2012}.
We opted to analyze the results that follow from our model, as they
should be valid for a similar energetic singlet and TT arrangement
in the absence of the aforementioned entropic mechanism. 

In organic solar cells TT yield is typically below $200\%$ because
the processes of singlet migration and charge separation are fast
enough to compete with $k_{SF}$ \cite{Congreve2013}. To consider
a similar situation, we assume that the singlet state quickly decays
to a charge-transfer state. In our model, this would correspond to
a scenario where there is a charge-acceptor next to each of the acene
molecules of the chain. For simplicity and for the purpose of showing
the possibilities of control of TT yield by polaritonic means, we
assume $k_{c}(\omega)=\delta_{\omega_{D,}\omega}k_{CT}$, ($\delta_{i,j}$
being the Kronecker delta function) where $k_{CT}$ is equal to the
bare pentacene rate $k_{SF}=17$ ps$^{-1}$. The form introduced for
$k_{c}(\omega)$ is approximately correct as long as the spectral
density describing the singlet-charge-transfer state is peaked around
$\omega_{D}$ and decays quickly with $\omega$. The aforementioned
$k_{CT}$ value is experimentally reasonable as it has been observed
in solar cells with a thin slab of SF material \cite{Congreve2013}. 

We use Eq. (\ref{eq:TT_yield}) to compute the $\epsilon_{TT}$ values
summarized in Table \ref{tab:param} in the presence of $k_{CT}$.
Our definition of yield is thus different from that obtained in steady-state,
but follows the spirit of many experiments that measure SF using time-domain
spectroscopy techniques \cite{Smith2010,Rao2010,Walker2013}. 

Turning now to the polariton-assisted SF case, the non-trivial dynamics
that emerge are due to differences in the density of states (DOS)
between the polariton and exciton manifolds, as well as to the photonic
character of each polariton state. These traits are encoded in the
prefactors $|c_{ph}^{\pm}|^{2}$, $|c_{ex}^{\pm}|^{2}$, $N$ and
$1/N$ in Eqs. (\ref{kin_up})-(\ref{kin_triplet}). Notably, in the
$N\gg1$ limit, the transfer rates from $P_{D}^{total}$ and $P_{TT}^{total}$
to the polariton manifold are largely suppressed. This is a consequence
of the large DOS of the former (which act like a population sink)
and the small DOS of the latter (which is spectrally isolated). The
reverse transfers are fast as they have single-molecule relaxation
scalings and correspond to going \emph{into} the population sink.
Similar findings are reported in Ref. \citen{delPino2015}, in the
context of the dynamics of molecular vibrations under the SC regime
and in \cite{Agranovich2003,Litinskaia2004} for exciton polaritons.
Therefore, once population reaches the dark and TT states, it is no
longer transferred back to the polariton manifolds, and the subsequent dynamics
is determined by transfer rates between dark and TT states. We stress,
however, that such asymmetry is approximate, as we
are ignoring the polariton bandwidth that emerges from the many photonic
modes hosted by the microcavity, which yields non-zero transfer rates
between to the aforementioned manifolds \cite{Litinskaia2004,Du2017}.

\begin{figure*}[t]
\includegraphics[width=0.8\paperwidth]{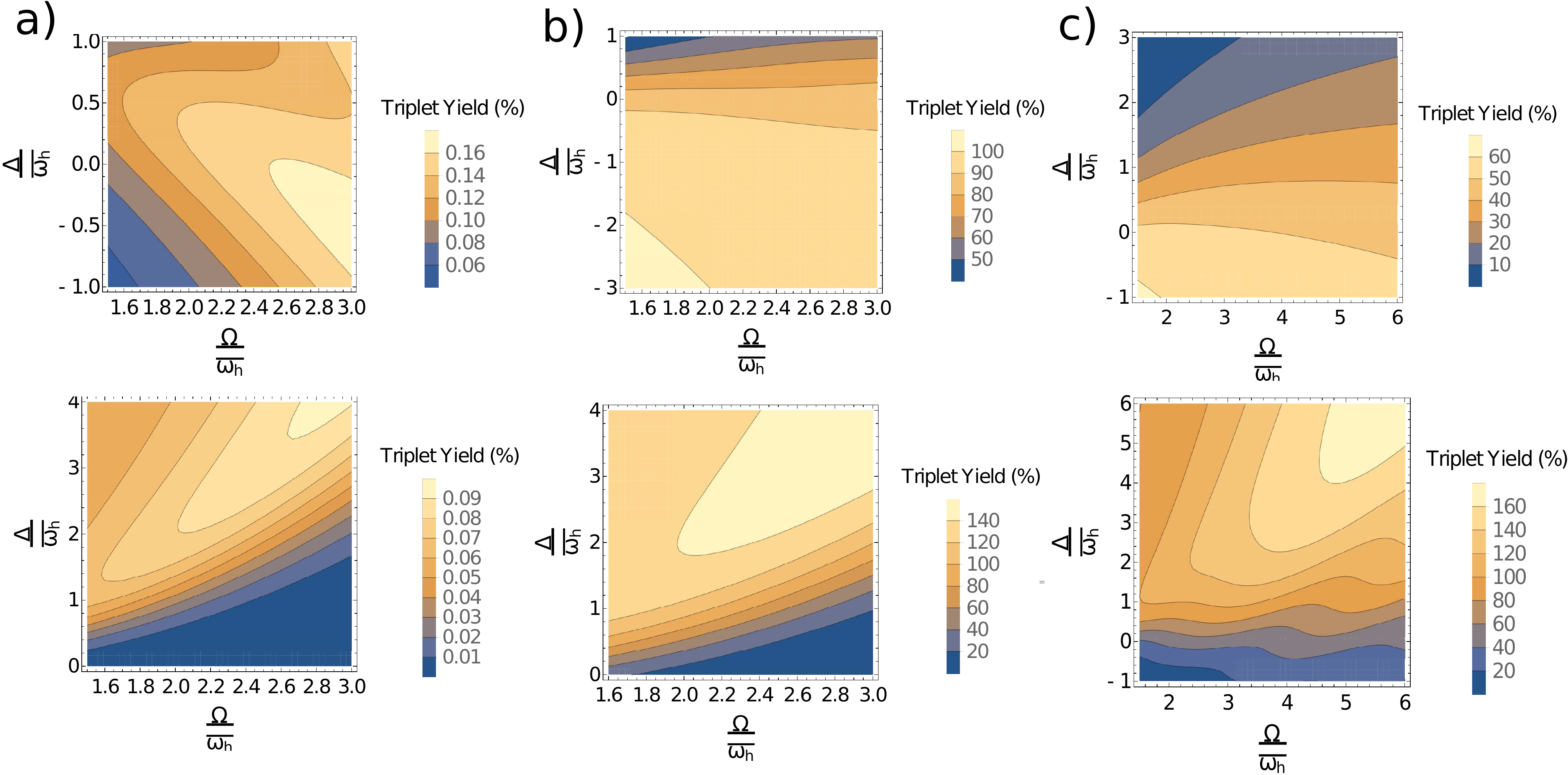}

\caption{TT yield as a function of $\Delta/\overline{\omega}_{h}=(\omega_{ph}-\omega_{e})/\overline{\omega}_{h}$
and $\Omega/\overline{\omega}_{h}$ featured by a) tetracene, b) pentacene
and c) hexacene. The upper (lower) plot considers initial conditions
with upper (lower) polariton pumping, \emph{i.e. }$P_{+}(0)=1$,$P_{a\protect\neq+}(0)=0$
($P_{-}(0)=1$, $P_{a\protect\neq-}(0)=0$) . \label{fig:TT_yield}}
\end{figure*}

We performed numerical simulations of the dynamics of the polariton-assisted
scenario by assuming two different initial conditions: pumping of the
upper polariton (UP) $\left(P_{+}(0)=1,P_{a\neq+}(0)\right)$ and
of the lower polariton (LP) $\left(P_{-}(0)=1,P_{a\neq0}(0)\right)$
for mentioned acenes. We denote $\tilde{\epsilon}_{TT}$ as the polariton-assisted
TT yields. They were calculated using the same criteria as in Eq.
(\ref{eq:TT_yield}). From a comparison of $\epsilon_{TT}$ (Table
\ref{tab:param}) and the values $\tilde{\epsilon}_{TT}(\Delta,\Omega)$
(where $\Delta=\omega_{ph}-\omega_{D}$ is the detuning between the
cavity photon and the singlet) in Fig. \ref{fig:TT_yield} we notice
an enhancement in the TT yield for hexacene, especially when the LP
is pumped. Under these conditions and high $\Delta$ values, the state
$|-\rangle$ is almost purely excitonic ($|c_{ex}^{-}|^{2}\approx1$),
and the rate of the channel associated to photon leakage is suppressed.
Moreover, $\omega_{-}$ becomes closer to resonance with the third
vibrational state of the TT manifold (with frequency $\overline{\omega}_{h}$)
for a given range of $\Omega$. Thus population transfer $|-\rangle\rightarrow\{|T_{n}T_{n+1}\rangle\}$
is faster than the bare SF, as the energetic barrier is lower in the
former. Finally, there is no competition between the previous transfer
process and the decay channel associated to $k_{c}$, as $k_{c}(\omega_{-})=0$,
under the assumptions of our model. The enhancement considering pumping
of the UP for the same molecule (Fig. \ref{fig:TT_yield} c), upper) is weaker since the fast rate of the
transfer $|+\rangle\rightarrow\{|d\rangle\}$ competes with the rate
of $|+\rangle\rightarrow\{|T_{n}T_{n+1}\rangle\}$. 

Pentacene shows a similar behavior: when $\Delta\approx-\overline{\omega}_{h}$
and the UP is pumped (Fig. \ref{fig:TT_yield} b), upper), then $|+\rangle$ is mainly excitonic and population
of the TT states is mainly determined by transfer from the dark state
manifold, since $\alpha(\omega_{+D})\gg\gamma(\omega_{TT,+})$, \emph{
i.e.} the population from the UP is quickly transferred
to dark states before transfer to TT states is carried out. Hence,
noting that $\gamma(\omega_{TT,D})\approx k_{SF}$, we recover (the
bare) pentacene yield $\epsilon_{TT}$. Notice however that for large detunings
$\tilde{\epsilon}_{TT}(\Delta\approx-3\overline{\omega}_{h},\Omega\approx1.6\overline{\omega}_{h})>\epsilon_{TT}$
because a phonon blockade prevents fast UP decay into dark states
and additionally the charge-transfer decay channel is suppressed ($k_{c}(\omega_{+})=0$).
On the other hand, $\tilde{\epsilon}_{TT}$ values are higher for pumping of the LP (Fig. \ref{fig:TT_yield} b), lower) as a result of a reduced decay rate $\alpha(\omega_{-D})$ to the dark states such that the predominant transfer process is from $|-\rangle$ to TT states.

Tetracene shows a distinct behavior in view of the bare energetic
arrangements of its singlet and TT states (see Fig. \ref{fig:decay_scheme}).
More concretely we have $|c_{ex}^{-}|^{2}\gamma(\omega_{TT,-}),|c_{ex}^{+}|^{2}\gamma(\omega_{TT,+}),\gamma(\omega_{TT,D})\ll\gamma(\omega_{D,TT})$
for most of the explored $(\Delta,\Omega)$ values. This translates
into a rapid depletion of the population of the TT manifold during
the considered timescale; which is a consequence of the energy of
the TT states lying above the dark state energy, in such a way that
the rate of population depletion of TT states towards the dark states
outcompetes the rate of the inverse process (in view of detailed balance)
as well as rates from the polariton manifold to the TT states. The
largest $\tilde{\epsilon}_{TT}$ values are reached when the UP is
pumped (Fig. \ref{fig:TT_yield} a), upper) and for parameters ($\Delta$,$\Omega$) which yield a predominantly
excitonic character to $|+\rangle$, and a sufficiently high rate
for the $|+\rangle\rightarrow\{|T_{n}T_{n+1}\rangle\}$ process, such
that it can better compete with the rates associated to $|+\rangle\rightarrow\{|d\rangle\}$
and $\{|T_{n}T_{n+1}\rangle\}\rightarrow\{|d\rangle\}$. Considering
the pumping of the LP (Fig. \ref{fig:TT_yield} a), lower), the maximal $\tilde{\epsilon}_{TT}$ values
are lower in view of $\omega_{TT}-\omega_{-}>0$, which greatly diminishes
the rate of the transfer $|-\rangle\rightarrow\{|T_{n}T_{n+1}\rangle\}$.

\begin{figure}
\includegraphics[scale=0.6]{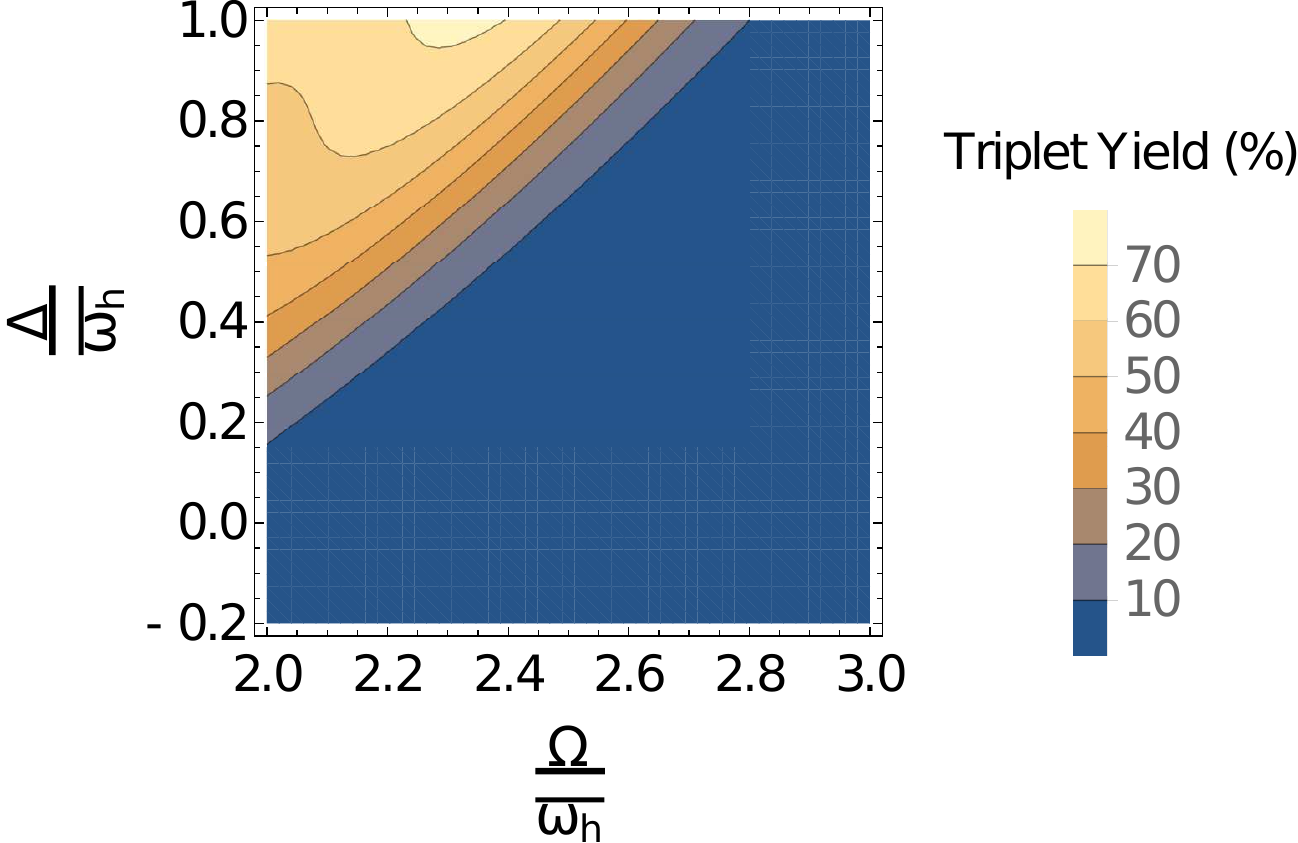}

\caption{TT yield as a function of $\Delta/\overline{\omega}_{h}=(\omega_{ph}-\omega_{e})/\overline{\omega}_{h}$
and $\Omega/\overline{\omega}_{h}$ featured by a molecule with $V_{S-TT}=44$
meV , $\Delta G=-630$ meV, and an outer reorganization energy of
$5$ meV. We consider  initial conditions with LP population ($P_{-}(0)=1$,
$P_{a\protect\neq-}(0)=0$). In this case the only competing decay
channel with SF is fluorescence.\label{fig:TTbadSF}}
\end{figure}

Until now, we have considered SF materials that already feature high
TT yields in the bare case if fast quenching mechanisms like charge
transfer are absent. We wonder if anything interesting remains if
the latter mechanisms are absent, i.e., if we set $k_{c}(\omega)=k_{f}=2.5\times10^{-4}\,\text{ps}^{-1}$
as an approximate fluorescence slow decay rate. To address this, we
consider a poor SF material with an outer sphere reorganization energy
of $5$ meV, while keeping the rest of the parameters as for hexacene.
This situation could correspond to hexacene in a solvent that significantly
increases the outer sphere SF energy barrier, such that $\epsilon_{TT}=18\%$.
Upon introducing a strongly coupled photonic mode, pumping the LP (Fig. \ref{fig:TTbadSF}) leads to $\tilde{\epsilon}_{TT}>\epsilon_{TT}$ for values $(\Delta,\Omega)$ that correspond to a transfer process $|-\rangle\rightarrow\{|T_{n}T_{n+1}\rangle\}$ whose rate competes with the rate of $|-\rangle\rightarrow\{|d\rangle\}$. 
This
occurs when $\omega_{-}$ is close to resonance with one of the high-frequency
vibronic states of the TT states, and $|-\rangle$ is predominantly
excitonic. Such a system could give a straightforward verification
of polariton assisted SF in the absence of a fast quenching process.

To summarize, in this letter, we have shown that when SF materials
are subjected to SC with a microcavity mode, the photonic leakage
of the resulting polariton states constitutes an important decay channel
that can decrease TT production, when compared to the bare case. However,
the rates associated with this competing decay channel can be tuned
by modifying the ratio $\Delta/\Omega$ (see Fig. \ref{fig:TT_yield}),
in such a way that the dynamics are dictated by the energy differences
and the DOS of the dressed states involved in SF. Given the large
DOS of the dark state manifold and TT states, the latter must lie
lower in energy with respect to the former so as to avoid population
leakage towards the dark state manifold and enhance TT yield. Remarkably,
while hexacene is a poor SF candidate in the bare case when quenchers
are present, it is the material which features the highest enhancement
under the proposed polaritonic approach and can even outcompete the
bare pentacene TT yield under the conditions considered in this work.
Similarly, we notice an increase of the pentacene TT yield, although
the improvements are modest in comparison with those obtained for
hexacene. Finally, we have also considered the putative scenario of
a SF material with low TT yield, where the SF rate competes with fluorescence.
In this case, our model predicts (for acene-like molecules) that $\Delta G\ll0$
substantial enhancement of TT yield can be achieved by polariton methods.

\section*{Autor Information}

\subsection*{Corresponding author}

{*}E-mail: joelyuen@ucsd.edu

\section*{Acknowledgments}

L.A.M.M is grateful for the support of the UC-Mexus CONACyT scholarship
for doctoral studies and with Jorge Campos-Gonz\'alez-Angulo for
useful discussions. R.F.R., M.D. and J.Y.Z. acknowledge support from
the NSF CAREER award CHE-1654732. S.K.C. acknowledges support from
the Canada Research Chairs program and NSERC RGPIN-2014-06129. L.A.M.M.,
R.F.R., M.D. and J.Y.Z. are thankful with UCSD for generous startup
funds. L.A.M.M. and J. Y. Z. acknowledge Ming Lee Tang, Michael Tauber,
Shane R. Yost and Troy Van Voorhis for helpful comments.



\bibliography{singlet_fission}


\pagebreak
\widetext
\begin{center}
\textbf{\large SI: Polariton-assisted singlet fission in acene aggregates}
\end{center}
\setcounter{equation}{0}
\setcounter{figure}{0}
\setcounter{table}{0}
\setcounter{page}{1}
\makeatletter
\renewcommand{\theequation}{S\arabic{equation}}
\renewcommand{\thefigure}{S\arabic{figure}}
\renewcommand{\bibnumfmt}[1]{[S#1]}
\renewcommand{\citenumfont}[1]{S#1}

\section*{Description of the open-quantum-systems model}

In this Supporting Information (SI), we provide the theoretical basis
for the kinetic model presented in the main text. To begin, consider
the explicit form of the Hamiltonian given in Eq (1) of the main text:
\begin{align*}
H & =\Big(\omega_{e}+\sum_{i}\omega_{i}\lambda_{S,i}^{2}\Big)\sum_{n}|n\rangle\langle n|+\sum_{n,i}\omega_{i}b_{n,i}^{\dagger}b_{n,i}+\sum_{n,i}|n\rangle\langle n|\omega_{i}\lambda_{S,i}(b_{n,i}+\text{h.c.})\\
 & +\omega_{ph}a^{\dagger}a+\sum_{n}g\big(a^{\dagger}|G\rangle\langle n|+\text{h.c.}\big)\\
 & +\sum_{n=0}^{N-1}\Big(\omega_{TT}+2\sum_{i}\omega_{i}\lambda_{T,i}^{2}\Big)|T_{n}T_{n+1}\rangle\langle T_{n}T_{n+1}|+\sum_{n=0}^{N-1}|T_{n}T_{n+1}\rangle\langle T_{n}T_{n+1}|\sum_{i}\omega_{i}\lambda_{T,i}(b_{n,i}+b_{n+1,i}+\text{h.c.})\\
 & +\frac{V_{TT-S}}{2}\sum_{n=0}^{N-1}\Big[\left(|T_{n}T_{n+1}\rangle+|T_{n-1}T_{n}\rangle\right)\langle n|+\text{h.c.}\Big].
\end{align*}

Assuming periodic boundary conditions $\left(|T_{N-1}T_{N}\rangle=|T_{N-1}T_{0}\rangle\right)$,
we introduce delocalized Fourier bases for the singlet excitons $|k\rangle=\frac{1}{\sqrt{N}}\sum_{n}e^{ikn}|n\rangle$
and vibrational degrees of freedom $b_{q,i}^{\dagger}=\frac{1}{\sqrt{N}}\sum_{n}e^{iqn}b_{n,i}^{\dagger}$,
where $k,q=\frac{2\pi m}{N},\,m=0,1,2,\dots,N-1$. We can then rewrite
the Hamiltonian above as

\begin{align}
H & =\sum_{k}(\omega_{e}+\sum_{i}\omega_{i}\lambda_{S,i}^{2})|k\rangle\langle k|+\sum_{q,i}\omega_{i}b_{q,i}^{\dagger}b_{q,i}+\frac{1}{\sqrt{N}}\sum_{q,k}\sum_{i}\omega_{i}\lambda_{S,i}\Big(|k\rangle\langle k+q|b_{q,i}^{\dagger}+\text{h.c.}\Big)\\
 & +\omega_{ph}a^{\dagger}a+\sqrt{N}g\Big(a^{\dagger}|G\rangle\langle k=0|+\text{h.c.}\Big)\\
 & +\sum_{n=0}^{N-1}\big(\omega_{TT}+2\sum_{i}\omega_{i}\lambda_{T,i}^{2}\big)|T_{n}T_{n+1}\rangle\langle T_{n}T_{n+1}|+\sum_{n=0}^{N-1}|T_{n}T_{n+1}\rangle\langle T_{n}T_{n+1}|\sum_{i}\lambda_{T,i}\omega_{i}(b_{n,i}+b_{n+1,i}+\text{h.c.})\\
 & +\frac{V_{TT-S}}{2}\sum_{n=0}^{N-1}\Big[\left(|T_{n}T_{n+1}\rangle+|T_{n}T_{n-1}\rangle\right)\langle n|+\text{h.c.}\Big],\nonumber 
\end{align}
 The introduction of the delocalized basis in the singlet electronic
manifold is the starting point to find approximate eigenstates for
$H$ in the strong-coupling (SC) regime, such that the rates of population
transfer between these eigenstates can be calculated by a perturbative
approach. Along these lines, we introduce a zeroth order Hamiltonian:
\begin{align}
H_{0} & =\sum_{k}\Big[\omega_{e}+\sum_{i}\omega_{i}\lambda_{S,i}^{2}+\frac{1}{\sqrt{N}}\sum_{i}\omega_{i}\lambda_{S,i}(b_{q=0,i}+b_{q=0,i}^{\dagger})\Big]|k\rangle\langle k|\nonumber \\
 & +\sum_{n',i}\omega_{i}b_{n',i}^{\dagger}b_{n',i}\Big(|G\rangle\langle G|+\sum_{k}|k\rangle\langle k|+\sum_{n=0}^{N-1}|T_{n}T_{n+1}\rangle\langle T_{n}T_{n+1}|\Big)\nonumber \\
 & +\omega_{ph}a^{\dagger}a+\sqrt{N}g\Big(a^{\dagger}|G\rangle\langle k=0|+\text{h.c.}\Big)\label{eq:zeroth_order}\\
 & +\sum_{n=0}^{N-1}\Big[\omega_{TT}+2\sum_{i}\omega_{i}\lambda_{T,i}^{2}+\sum_{i}\lambda_{T,i}\omega_{i}(b_{n,i}+b_{n,i}^{\dagger}+b_{n+1,i}+b_{n+1,i}^{\dagger})\Big]|T_{n}T_{n+1}\rangle\langle T_{n}T_{n+1}|\nonumber \\
 & =\sum_{n,i}\omega_{i}b_{n,i}^{\dagger}b_{n,i}|G\rangle\langle G|\nonumber \\
 & +\sum_{k}\Bigg[\omega_{e}+\sum_{i}\omega_{i}\tilde{b}_{q=0,i}^{\dagger}\tilde{b}_{q=0,i}+\sum_{q\neq0,i}\omega_{i}b_{q,i}^{\dagger}b_{q,i}+\Big(1-\frac{1}{N}\Big)\sum_{i}\omega_{i}\lambda_{S,i}^{2}\Bigg]|k\rangle\langle k|+\omega_{ph}a^{\dagger}a+\sqrt{N}g\Big(a^{\dagger}|G\rangle\langle k=0|+\text{h.c.}\Big)\label{eq:zeroth_order_2}\\
 & +\sum_{n=0}^{N-1}\left\{ \omega_{TT}+\sum_{i}\Bigg[\omega_{i}\Big(\overline{b}_{n,i}^{\dagger}\overline{b}_{n,i}+\overline{b}_{n+1,i}^{\dagger}\overline{b}_{n+1,i}\Big)+\sum_{n'\neq n,n+1}\omega_{i}b_{n',i}^{\dagger}b_{n',i}\Bigg]\right\} |T_{n}T_{n+1}\rangle\langle T_{n}T_{n+1}|\nonumber 
\end{align}
where we have used, $\sum_{n,i}\omega_{i}b_{n,i}^{\dagger}b_{n,i}=\sum_{q,i}\omega_{i}b_{q,i}^{\dagger}b_{q,i}=\sum_{n',i}\omega_{i}b_{n',i}^{\dagger}b_{n',i}\Big(|G\rangle\langle G|+\sum_{k}|k\rangle\langle k|+\sum_{n=0}^{N-1}|T_{n}T_{n+1}\rangle\langle T_{n}T_{n+1}|\Big)$,
and the definitions $\tilde{b}_{q=0,i}^{\dagger}=b_{q=0,i}^{\dagger}+\lambda_{S,i}/\sqrt{N}$
, $\overline{b}_{n,i}^{\dagger}=b_{n,i}^{\dagger}+\lambda_{T,i}$
([R1,R2]). Accordingly, the perturbation is
defined as $V=H-H_{0}$ that is

\begin{subequations}\label{pert}

\begin{align}
V & =\frac{1}{\sqrt{N}}\sum_{q\neq0,k}\sum_{i}\omega_{i}\lambda_{S,i}\Big(|k\rangle\langle k+q|b_{q,i}^{\dagger}+\text{h.c.}\Big)\label{first_pert}\\
 & +\frac{V_{TT-S}}{2}\sum_{n=0}^{N-1}\bigg[\left(|T_{n}T_{n+1}\rangle+|T_{n}T_{n-1}\rangle\right)\langle n|+\text{h.c.}\bigg]\label{tt_pert}\\
 & =\frac{1}{\sqrt{N}}\sum_{q\neq0,k}\sum_{i}\bigg[\omega_{i}\lambda_{S,i}|k\rangle\langle k+q|\big(b_{q,i}^{\dagger}+b_{-q,i}\big)\bigg]\nonumber \\
 & +\frac{V_{TT-S}}{2}\sum_{n=0}^{N-1}\bigg[\left(|T_{n}T_{n+1}\rangle+|T_{n}T_{n-1}\rangle\right)\langle n|+\text{h.c.}\bigg],\nonumber 
\end{align}

\end{subequations}\noindent which accounts for the scattering of
the singlet electronic states due to phonons (Eq. (\ref{first_pert}))
and the coupling of singlet to TT states (Eq. (\ref{tt_pert})). The
diagonalization of Eq. (\ref{eq:zeroth_order_2}) is carried out as
follows: since our rate calculations are performed in the linear response
regime with respect to an external pump, we consider ony zero- and
one-photon-dressed states. In the limit $N\gg1$, $\tilde{b}_{q=0,i}^{\dagger}\approx b_{q=0,i}$ [R1]
for all $i$ and $\frac{1}{N}\approx0$ which results in the diagonal
Hamiltonian

\begin{align}
H_{0} & \approx\sum_{n,i}\omega_{i}b_{n,i}^{\dagger}b_{n,i}|G\rangle\langle G|\label{eq:zeroth_eigenstates}\\
 & +\big(\omega_{+}+\sum_{i,q}\omega_{i}b_{q,i}^{\dagger}b_{q,i}\big)|+\rangle\langle+|+\big(\omega_{-}+\sum_{i,q}\omega_{i}b_{q,i}^{\dagger}b_{q,i}\big)|-\rangle\langle-|\nonumber \\
 & +\sum_{d}\big(\omega_{e}+\sum_{i,q}\omega_{i}b_{q,i}^{\dagger}b_{q,i}+\sum_{i}\omega_{i}\lambda_{S,i}^{2}\big)|d\rangle\langle d|\nonumber \\
 & +\sum_{n=0}^{N-2}\Bigg[\omega_{TT}+\sum_{i}\omega_{i}\Big(\overline{b}_{n,i}^{\dagger}\overline{b}_{n,i}+\overline{b}_{n+1,i}^{\dagger}\overline{b}_{n+1,i}\Big)+\sum_{n'\neq n,n+1}\omega_{i}b_{n',i}^{\dagger}b_{n',i}\Bigg]|T_{n}T_{n+1}\rangle\langle T_{n}T_{n+1}|\nonumber \\
 & =H_{S}^{vib}|G;0\rangle\langle G;0|\label{zeroth_eigenstates_2}\\
 & +\big(\omega_{+}+H_{S}^{vib}\big)|+\rangle\langle+|+\big(\omega_{-}+H_{S}^{vib}\big)|-\rangle\langle-|\nonumber \\
 & +\sum_{d}\big(\omega_{D}+H_{S}^{vib}\big)|d\rangle\langle d|\nonumber \\
 & +\sum_{n=0}^{N-2}\Bigg[\omega_{TT}+H_{T_{n}T_{n+1}}^{vib}\Bigg]|T_{n}T_{n+1}\rangle\langle T_{n}T_{n+1}|\nonumber 
\end{align}
where $|G;0\rangle=|G\rangle\otimes|0_{ph}\rangle$ ($|n_{ph}\rangle$
is the state with $n$ photons) the vibrational Hamiltonians are $H_{S}^{vib}=\sum_{n,i}\omega_{i}b_{n,i}^{\dagger}b_{n,i}$
and $H_{T_{n}T_{n+1}}^{vib}=\sum_{i}\Bigg[\omega_{i}\Big(\overline{b}_{n,i}^{\dagger}\overline{b}_{n,i}+\overline{b}_{n+1,i}^{\dagger}\overline{b}_{n+1,i}\Big)+\sum_{n'\neq n,n+1}\omega_{i}b_{n',i}^{\dagger}b_{n',i}\Bigg]$,
and we have invoked the convenient notation $|d\rangle=|k\neq0\rangle$,
$d=\frac{2\pi m}{N}$ for $m=1,\cdots,N-1$ to denote the dark states.
The upper and lower polariton eigenstates are given by

\begin{align}
|+\rangle & =\cos\theta|G\rangle\otimes|1_{ph}\rangle+\sin\theta|k=0\rangle\otimes|0_{ph}\rangle,\nonumber \\
|-\rangle & =-\sin\theta|G\rangle\otimes|1_{ph}\rangle+\cos\theta|k=0\rangle\otimes|0_{ph}\rangle,\label{eq:fractions}
\end{align}
with their corresponding eigenenergies given in Eq. (7) of the main
text. The factors $\cos\theta$ and $\sin\theta$ account for the
degree of mixing of the considered photon-dressed electronic states,
where $\text{tan}2\theta=\frac{2\sqrt{N}g}{\omega_{e}+\sum_{i}\omega_{i}\lambda_{S,i}^{2}-\omega_{ph}}$.
Notice that the corresponding vibrational Hamiltonians for $|G\rangle$,
$|+\rangle$, $|-\rangle$, and $\{|d\rangle\}$ are all the same
$\left(H_{S}^{vib}\right)$ in this zeroth-order treatment. Here,
we note that the energy offset of the dark states is at $\omega_{D}=\omega_{e}+\sum_{i}\omega_{i}\lambda_{S,i}^{2}$.

\textbf{~}

\section*{Calculation of kinetic rates}

We now describe the reduced dynamics of the polaritonic and excitonic
states (namely $|\pm\rangle$, $\{|d\rangle\}$, $\{|T_{n}T_{n+1}\rangle\}$)
due to the perturbation $V$. Since we are only interested in the
evolution of populations, we derive a Pauli master equation using
the Redfield formalism under the secular and Markov approximations
[R3], 
\[
\frac{dP_{a}(t)}{dt}=\sum_{b\neq a}\Big(k_{b\rightarrow a}P_{b}-k_{a\to b}P_{a}\Big)
\]
where $P_{a}(t)$, $a\in\{+,-,d,T_{n}T_{n+1}\}$, is the population
in state $a$. The population-transfer rate from state $b$ to state
$a$ is given by [R3],
\begin{equation}
k_{b\rightarrow a}=\int_{-\infty}^{\infty}d\tau\langle V_{ba}(\tau)V_{ab}(0)\rangle,\label{eq:fermis}
\end{equation}
which is Fermi's golden rule in the form a correlation function Fourier
transform; here, $V_{ba}(\tau)=\langle b|e^{iH_{0}\tau}Ve^{-iH_{0}\tau}|a\rangle=e^{-i\omega_{ab}\tau}\langle b|e^{iH_{b}^{vib}\tau}Ve^{-iH_{a}^{vib}\tau}|a\rangle$,
where we are using $\omega_{D}$, $\omega_{T_{n}T_{n+1}}=\omega_{TT}$
as the bare energies of the dark and TT states respectively, and $\omega_{ab}=\omega_{a}-\omega_{b}$.
Here, $\langle O\rangle=\text{Tr}_{vib}\{O\rho_{eq}^{vib}\}$ denotes
a vibrational trace over the thermal ensemble of phonons governed
by $H_{S}^{vib}$ at inverse temperature $\beta$, $\rho_{eq}^{vib}=\frac{e^{-\beta H_{S}^{vib}}}{\text{Tr}_{v}(e^{-\beta H_{S}^{vib}})}$.

Upon insertion of Eq. (\ref{pert}) into Eq. (\ref{eq:fermis}), the
population transfer between a state $|a\rangle\in\mathcal{S}$, $\mathcal{S}=\{|+\rangle,|-\rangle,|k\neq0\rangle\}$
and a state $|b\rangle\in\mathcal{TT}$, $\mathcal{TT}=\{|T_{n}T_{n+1}\rangle\}$
results in a Marcus-type expression [R3]. For instance,
the rate from $|+\rangle$ to $|T_{0}T_{1}\rangle$ is given by 
\begin{align}
k_{+\rightarrow T_{0}T_{1}} & =\int_{-\infty}^{\infty}\langle V_{+,T_{0}T_{1}}(\tau)V_{T_{0}T_{1},+}(0)\rangle d\tau\nonumber \\
 & =\frac{|V_{TT-S}|^{2}f_{+}}{N}\int_{-\infty}^{\infty}d\tau\left\langle \exp\Big(iH_{S}^{vib}\tau\Big)\exp\Big(-iH_{T_{0}T_{1}}^{vib}\tau\Big)\right\rangle e^{-i\omega_{TT+}\tau}\\
 & =\frac{|V_{TT-S}|^{2}f_{+}}{N}\int_{-\infty}^{\infty}d\tau\left\langle \exp\Bigg[i\Big(\sum_{i,n}\omega_{i}b_{n,i}^{\dagger}b_{n,i}\Big)\tau\Bigg]\exp\Bigg\{-i\Big[\sum_{i}\omega_{i}\big(\overline{b}_{0,i}^{\dagger}\overline{b}_{0,i}+\overline{b}_{1,i}^{\dagger}\overline{b}_{1,i}\big)\right.\nonumber \\
 & \left.+\sum_{i,m\neq0,1}\omega_{i}b_{m,i}^{\dagger}b_{m,i}\Big]\tau\Bigg\}\right\rangle e^{-i\omega_{TT,+}\tau},\label{eq:rate_UP_TT}
\end{align}
where we use the notation $f_{i}=\langle i|\mathcal{I}_{exc}^{S}|i\rangle$
($\mathcal{I}_{exc}^{S}=\sum_{n}|n\rangle\langle n|$) for the singlet-exciton
fraction in state $i$; here, $f_{+}=|\sin\theta|^{2}$. The $1/N$
prefactor accounts for the portion of the delocalized polariton state
which is in sites 0 and 1 (see Eq. (\ref{eq:fractions})) and can
thus undergo SF at these positions. To evaluate Eq. (\ref{eq:rate_UP_TT}),
we now introduce the gap Hamiltonian $H_{T_{0}T_{1},S}^{vib}=H_{T_{0}T_{1}}^{vib}-H_{S}^{vib}$
[R3], 
\begin{equation}
k_{+\rightarrow T_{0}T_{1}}=\frac{|V_{TT-S}|^{2}f_{+}}{N}\hat{T}\int_{-\infty}^{\infty}d\tau\left\langle \exp\Big[-i\int_{0}^{\tau}H_{T_{0}T_{1},S}^{vib}(t')dt'\Big]\right\rangle e^{-i\omega_{TT,+}\tau},\label{rate_UP_TT_2}
\end{equation}
where $\hat{T}$ is the time-ordering operator and 
\begin{align*}
H_{T_{0}T_{1},S}^{vib}(t') & =\exp\Big(iH_{S}^{vib}t'\Big)H_{T_{0}T_{1},S}^{vib}\exp\Big(-iH_{S}^{vib}t'\Big)\\
 & =2\sum_{i}\lambda_{i,T}^{2}\omega_{i}+\sum_{i}\lambda_{i,T}\omega_{i}(b_{0,i}^{\dagger}e^{i\omega_{i}t'}+b_{0,i}e^{-i\omega_{i}t'}+b_{1,i}^{\dagger}e^{i\omega_{i}t'}+b_{1,i}e^{-i\omega_{i}t'}).
\end{align*}
We approximate Eq. (\ref{rate_UP_TT_2}) with a cumulant expansion
\begin{align}
\hat{T}\left\langle \exp\Big[-i\int_{0}^{t}d\tau H_{T_{0}T_{1},S}^{vib}(\tau)\Big]\right\rangle  & \approx\exp\Bigg\{-i\int_{0}^{t}d\tau\langle H_{T_{0}T_{1},S}^{vib}(\tau)\rangle\label{correlation}\\
 & +(-i)^{2}\int_{0}^{t}d\tau_{2}\int_{0}^{\tau_{2}}d\tau_{1}\Big[\langle H_{T_{0}T_{1},S}^{vib}(\tau_{2})H_{T_{0}T_{1},S}^{vib}(\tau_{1})\rangle-\langle H_{T_{0}T_{1},S}^{vib}(\tau_{2})\rangle\langle H_{T_{0}T_{1},S}^{vib}(\tau_{1})\rangle\Big]\Bigg\},\nonumber 
\end{align}
where $\langle H_{T_{0}T_{1},S}^{vib}(\tau)\rangle=2\sum_{i}\lambda_{i,T}^{2}\omega_{i}$
and
\begin{align}
 & \int_{0}^{t}d\tau_{2}\int_{0}^{\tau_{2}}d\tau\bigg[\langle H_{T_{0}T_{1}S}^{vib}(\tau)H_{T_{0}T_{1}S}^{vib}(0)\rangle-\langle H_{T_{0}T_{1}S}^{vib}(\tau)\rangle\langle H_{T_{0}T_{1}S}^{vib}(0)\rangle\bigg]\label{lineshape}\\
= & \int_{0}^{t}d\tau_{2}\int_{0}^{\tau_{2}}d\tau\sum_{i}\bigg[\langle\lambda_{i,T}^{2}\omega_{i}^{2}\big(b_{0,i}e^{-i\omega_{i}\tau}+b_{0,i}^{\dagger}e^{i\omega_{i}\tau}+b_{1,i}e^{-i\omega_{i}\tau}+b_{1,i}^{\dagger}e^{i\omega_{i}\tau}\big)\nonumber \\
 & \times\big(b_{0,i}+b_{0,i}^{\dagger}+b_{1,i}+b_{1,i}^{\dagger}\big)\rangle\bigg]\nonumber \\
= & -\sum_{i}2\lambda_{i,T}^{2}\big[\big(\overline{n}_{i}(\omega_{i})+1\big)\big(e^{-i\omega_{i}t}-1\big)+\overline{n}_{i}(\omega_{i})\big(e^{i\omega_{i}t}-1\big)\big]-2i\sum_{j}\lambda_{j,T}^{2}\omega_{j}t\nonumber \\
= & \underbrace{-2\sum_{i}\lambda_{i,T}^{2}\big[\coth\big(\beta\omega_{i}/2\big)\big(\cos\omega_{i}t-1\big)-i\sin\omega_{i}t\big]}_{=G(t)}-2i\sum_{j}\lambda_{j,T}^{2}\omega_{j}t.\nonumber 
\end{align}
In Eq. (\ref{lineshape}), we have assumed a thermalized vibrational
bath and independent localized phonon modes, $\overline{n}_{i}(\omega_{i})=\frac{1}{e^{\beta\omega_{i}}-1}$
is the average bosonic occupation number of the $i$-th vibrational
mode. Using Eqs. (\ref{correlation}) and (\ref{lineshape}) altogether
yields,
\begin{equation}
k_{+\rightarrow T_{0}T_{1}}=\frac{|V_{TT-S}|^{2}f_{+}}{N}\int_{-\infty}^{\infty}d\tau e^{-i\big(\omega_{TT}-\omega_{+}\big)\tau}e^{-G(\tau)}.\label{eq:integral}
\end{equation}
The integral in Eq. (\ref{eq:integral}) can be simplified [R4]
by introducing a separation of low and high vibrational frequency
bath modes $G(t)\approx G_{l}(t)+G_{h}(t)$, where
\begin{align*}
G_{i}(t) & =-2\overline{\lambda}_{T,i}^{2}\big[\coth\big(\beta\overline{\omega}_{i}/2\big)\big(\cos\overline{\omega}_{i}t-1\big)-i\sin\overline{\omega}_{i}t\big].
\end{align*}
Here, $\overline{\omega}_{h}$ ($\overline{\omega}_{l}$) corresponds
to an effective high (low) frequency that satisfies $\overline{\omega}_{h}\gg1/\beta$
($\overline{\omega}_{h}\ll1/\beta$), while $2\overline{\lambda}_{T,h}^{2}$
($2\overline{\lambda}_{T,l}^{2}$) is an effective Huang-Rhys factor
for the bath mode with frequency $\overline{\omega}_{h}$ ($\overline{\omega}_{l}$).
Under this approach we obtain

\begin{subequations}\label{eq:FC_marcus}

\begin{align}
k_{+\rightarrow T_{0}T_{1}} & =\frac{1}{N}f_{+}\gamma_{d}(\omega_{TT}-\omega_{+}),\label{Marcus}\\
\gamma_{d}(\omega_{TT,+}) & =\sum_{n=0}^{\infty}\gamma^{(n)}(\omega_{TT}-\omega_{+}),\\
\gamma^{(n)}(\omega_{TT,+}) & =|V_{TT-S}|^{2}\sqrt{\frac{\pi\beta}{R_{l}}}\underbrace{e^{-2\overline{\lambda}_{T,h}^{2}}\frac{(2\overline{\lambda}_{T,h}^{2})^{n}}{n!}}_{\text{Frank-Condon (FC) factor}}\exp\Bigg[-\frac{\beta(R_{l}+\omega_{TT}+n\overline{\omega}_{h}-\omega_{+})^{2}}{4R_{l}}\Bigg],\label{eq:MarcusFC}
\end{align}

\end{subequations}\noindent where $R_{l}=2\overline{\omega}_{l}\overline{\lambda}_{T,l}^{2}$
is the reorganization energy of the low-frequency vibrational mode
and $\gamma^{(n)}(\omega_{ba})$ can be interpreted as a single-molecule
rate from the lowest-energy vibronic state of $|a\rangle$ to the
vibronic state of $|b\rangle$ with $n$ phonons in the high-frequency
vibrational mode. Notice that Eq. (\ref{Marcus}) does not feature
an energy scale associated to the reorganization energy of the singlet,
as a result of polaron decoupling [R1]. 

The rates $k_{-\rightarrow T_{0}T_{1}}$ , $k_{d\rightarrow T_{0}T_{1}}$
can be calculated in analogous ways giving
\begin{align}
k_{-\rightarrow T_{0}T_{1}} & =\frac{1}{N}f_{-}\gamma_{d}(\omega_{TT}-\omega_{-}),\label{eq:FC_Marcus_2}\\
k_{d\rightarrow T_{0}T_{1}} & =\frac{1}{N}\gamma_{d}(\omega_{TT}-\omega_{e}-\sum_{i}\omega_{i}\lambda_{S,i}^{2}).\label{eq:FC_Marcus_3}
\end{align}
The rates of backward transfer can be calculated by invoking detailed
balance, i.e., $k_{a\to b}=e^{\beta\omega_{ab}}k_{b\to a}$. Due to
similarity in chemical structure, we assume that the values for $\overline{\omega}_{h}=174$
meV and $\overline{\lambda}_{T,h}=1.6$ are independent of the identity
of the acene. The latter was estimated as $\overline{\lambda}_{T,h}=\overline{\lambda}_{S-TT,h}+\overline{\lambda}_{S,h}$,
where $\overline{\lambda}_{S-TT,h}$ is the square root of the Huang-Rhys
factor for the $|n\rangle\rightarrow|T_{n}T_{n+1}\rangle$ transition
[R5] and $\overline{\lambda}_{S,h}$ is the analogous
quantity for the $|G\rangle\rightarrow|n\rangle$ transition. Since
there is no available data in the literature to estimate $R_{l}$,
we opted to use the low-frequency reorganization energy of the transition
$|n\rangle\rightarrow|T_{n}T_{n+1}\rangle$ for hexacene, and we set
$R_{l}=100$ meV [R5] for all the materials considered.
On the other hand the parameters $\Delta G=\omega_{TT}-\omega_{e}$,
and $V_{TT-S}/2$ are treated as material dependent; they are taken
from Ref. [R6] and are given in the main text.

Marcus rates for the cavity-free case are approximately one order
of magnitude larger than the experimental ones when $V_{TT-S}$ is
large, \emph{i.e.} in the adiabatic limit. In the section below we
introduce corrections to these rates to make a connection with the
Bixon-Jortner equation [R7] which correctly interpolates
between the diabatic and adiabatic limits. 

We now consider the calculation of transfer rates between states $|a\rangle,|b\rangle\in\mathcal{S}$.
As an example, the transfer rate between the upper polariton and one
dark state $|d\rangle$ is given by
\begin{align}
k_{+\rightarrow d} & =\int_{-\infty}^{\infty}\langle V_{+d}(\tau)V_{d+}(0)\rangle d\tau\nonumber \\
 & =\frac{1}{N}f_{+}\sum_{i,j}\omega_{i}\lambda_{S,i}\omega_{j}\lambda_{S,j}\int_{-\infty}^{\infty}\left\langle e^{iH_{S}^{vib}\tau}(b_{q=d,i}+b_{q=-d,i}^{\dagger})e^{-iH_{d}^{vib}\tau}(b_{q=d,j}^{\dagger}+b_{q=-d,j})\rho_{eq}^{vib}\right\rangle \nonumber \\
 & \times e^{-i\Big[\sum_{i'}\omega_{i'}\lambda_{S,i'}^{2}+\omega_{e}-\omega_{+}\Big]\tau}d\tau\nonumber \\
 & =\frac{1}{N}f_{+}\sum_{i}|\omega_{i}\lambda_{S,i}|^{2}\int_{-\infty}^{\infty}\langle(b_{q=d,i}e^{-i\omega_{i}\tau}+b_{q=-d,i}^{\dagger}e^{i\omega_{i}\tau})(b_{q=d,i}^{\dagger}+b_{q=-d,i})\rangle e^{-i\big(\sum_{i'}\omega_{i'}\lambda_{S,i'}^{2}+\omega_{e}-\omega_{+}\big)\tau}d\tau\label{eq:dark_transfer}\\
 & =\frac{1}{N}f_{+}\sum_{i}|\omega_{i}\lambda_{S,i}|^{2}\int_{-\infty}^{\infty}\langle b_{q=d,i}b_{q=d,i}^{\dagger}e^{-i\omega_{i}\tau}+b_{q=-d,i}^{\dagger}b_{q=-d,i}e^{i\omega_{i}\tau}\rangle e^{-i\big(\sum_{i'}\omega_{i'}\lambda_{S,i'}^{2}+\omega_{e}-\omega_{+}\big)\tau}d\tau,\nonumber 
\end{align}
where we used the fact that the phonons are uncorrelated to simplify
the double summation. Notice that in our model the energy of the phonon
modes is independent of their momentum (in fact, the calculation using
a localized phonon basis would render the same answer [R8]).
Identifying $\langle b_{d,i}b_{d,i}^{\dagger}e^{-i\omega_{i}\tau}+b_{-d,i}^{\dagger}b_{-d,i}e^{i\omega_{i}\tau}\rangle=C(\tau,0)$
as a bath correlation function [R3],

\[
C(\tau,0)=\int_{0}^{\infty}J(\omega)\big[\big(\overline{n}(\omega)+1\big)e^{-i\omega\tau}+\overline{n}(\omega)e^{i\omega\tau}\big]d\omega,
\]
where $J(\omega)=2\sum_{i}|\omega_{i}\lambda_{S,i}|^{2}\delta(\omega-\omega_{i})$
is the bath spectral density, we obtain
\begin{align*}
k_{+\rightarrow d} & =\frac{f_{+}}{N}\alpha\Bigg(\omega_{+}-\sum_{i}\omega_{i}\lambda_{S,i}^{2}-\omega_{e}\Bigg),
\end{align*}
where $\alpha(\omega)=\pi J(\omega)[\overline{n}(\omega)+1]$ for
$\omega\geq0$ and $\alpha(\omega)=\pi J(\omega)\overline{n}(\omega)$
for $\omega<0$. In our calculations, we use an Ohmic spectral density
with a Lorentzian frequency cutoff: $J(\omega)=2R\Omega\omega\frac{1}{\omega^{2}+\Omega^{2}}$
with the parameters $R=50$ meV and $\Omega=180$ meV taken from Ref.
[R9] and are assumed to be the same for the acenes
considered. The same approach can be used to compute the rates of
relaxation $k_{+\rightarrow-}$, $k_{d\rightarrow-}$, and the backward
rates inferred from the detailed balance condition, as explained before.
It is noteworthy that our approach predicts rates $k_{+\rightarrow-}=k_{-\rightarrow+}=0$.
This follows from the perturbation (see Eq. (\ref{pert})), which
does not couple the $|+\rangle$ and $|-\rangle$ states. 

\textbf{Adiabatic corrections to Marcus equations.\textemdash{} }To
make a stronger connection with experimental circumstances, we will
now consider an \emph{ad-hoc} way to include corrections to the Marcus-type
treatment above, so that our predictions remain valid throughout the
diabatic (small $V_{TT-S}$) and adiabatic (large $V_{TT-S}$) regimes.
To do so, we renormalize the rates in Eq. (\ref{eq:MarcusFC}) $\gamma^{(n)}(\omega_{ij})\rightarrow\gamma^{(n)}(\omega_{ij})e^{-\Gamma^{(n)}}$
[R3] where
\begin{align*}
\Gamma^{(n)} & =\frac{2\pi}{v}\frac{|V_{TT-S}|^{2}}{|F_{S}-F_{T}|}\underbrace{e^{-2\overline{\lambda}_{T,h}^{2}}\frac{(2\overline{\lambda}_{T,h}^{2})^{n}}{n!}}_{\text{FC factor}}
\end{align*}
is a FC-corrected Massey parameter with $v$ being a characteristic
velocity of the (slow) low-frequency coordinate; $|F_{S}-F_{TT}|$
is the corresponding absolute difference in slopes of these potential
energy surfaces (PESs) at their crossing point. The factor $e^{-\Gamma^{(n)}}$
accounts for the probability the system will remain in the initial
PES after times much longer than $\frac{2\pi}{\overline{\omega}_{l}}$.
Taylor expanding the correction factor, we obtain,
\[
\frac{\gamma^{(n)}(\omega_{ij})}{e^{\Gamma^{(n)}}}\approx\frac{\gamma^{(n)}(\omega_{ij})}{1+\Gamma^{(n)}}.
\]
To define $\Gamma^{(n)}$ in terms of parameters of our model we have
$|F_{S}-F_{TT}|=2\left|\frac{R_{l}}{\sqrt{\frac{2}{m\overline{\omega}_{l}}}\overline{\lambda_{T,l}}}\right|$
where $m$ and $R_{l}$ are the mass and the reorganization energy
of the effective low-frequency vibrational mode introduced previously.
To obtain an estimate for $v$, we consider the root-mean square of
the velocity $\langle v^{2}\rangle^{1/2}$ around the minimum of the
singlet PES which in the high temperature limit can be calculated
as [R7], 
\[
v=\langle v^{2}\rangle^{1/2}\approx\frac{\overline{\lambda_{T,l}}}{\tau_{d}}\sqrt{\frac{2}{m\beta\overline{\omega}_{l}^{2}}}\equiv\frac{\overline{\lambda_{T,l}}}{\tau_{nad}},
\]
where $\tau_{d}$ is the decay time of the position autocorrelation
function for the low-frequency vibrational mode, and we have accordingly
introduced the non-adiabatic timescale $\tau_{nad}$. By using the
previous set of approximations, we heuristically obtain the Bixon-Jortner
equation [R7], which has been shown to predict
SF rates close to experimental values [R6]. Collecting
the results above, Eqs. (\ref{eq:FC_marcus})\textendash (\ref{eq:FC_Marcus_3})
become:
\begin{align}
k_{i\rightarrow j} & =\frac{f_{i}}{N}\sum_{n=0}^{\infty}\frac{\gamma^{(n)}(\omega_{ij})}{1+\frac{4\pi|V_{S-T}|^{2}}{\hbar R_{l}}\tau_{nad}e^{-2\overline{\lambda}_{T,h}^{2}}\frac{(2\overline{\lambda}_{T,h}^{2})^{n}}{n!}}.\label{eq:BJ_eqn}\\
 & =\frac{f_{i}}{N}\sum_{n=0}^{\infty}\gamma^{(n)}(\omega_{ij})\Gamma_{ad}^{(n)},\\
 & =\frac{f_{i}}{N}\gamma(\omega_{ij})
\end{align}
assuming $|i\rangle\in\mathcal{S}$ and $|j\rangle\in\mathcal{TT}$.
Notice that the $\gamma(\omega)$ rates introduced above are the ones
used in the main text.

\section*{Final kinetic equations}

With the rates calculated above we construct the following kinetic
model for SF in the SC regime:

\begin{align*}
\partial_{t}P_{\pm} & =-f_{\pm}\frac{\alpha(\omega_{\pm D})}{N}(N-1)P_{\pm}+f_{\pm}\sum_{d}\frac{\alpha(\omega_{D\pm})}{N}P_{d}-f_{\pm}\frac{\gamma(\omega_{TT,\pm})}{N}(N)P_{\pm}\\
 & +f_{\pm}\sum_{n=0}^{N-2}\frac{\gamma(\omega_{\pm,TT})}{N}P_{T_{n}T_{n+1}}-(f_{\pm}k_{c}(\omega_{\pm})+f_{\pm}^{p}k_{phot})P_{\pm}
\end{align*}

\begin{align*}
\partial_{t}P_{d} & =f_{+}\frac{\alpha(\omega_{+D})}{N}P_{+}-f_{+}\frac{\alpha(\omega_{D+})}{N}P_{d}+f_{-}\frac{\alpha(\omega_{-D})}{N}P_{-}-f_{-}\frac{\alpha(\omega_{D-})}{N}P_{d}-\frac{\alpha (0)}{N}(N-2)P_{d}\\
 & +\frac{\alpha (0)}{N}\sum_{d\neq d'}P_{d'd'}-\frac{\gamma(\omega_{TT,D})}{N}(N)P_{d}+\sum_{n}\frac{\gamma(\omega_{D,TT})}{N}\rho_{T_{n}T_{n+1}}-k_{c}(\omega_{D})\rho_{dd}
\end{align*}

\begin{align*}
\partial_{t}P_{T_{n}T_{n+1}} & =f_{+}\frac{\gamma(\omega_{TT,+})}{N}P_{+}-f_{+}\frac{\gamma(\omega_{+,TT})}{N}P_{T_{n}T_{n+1}}-\frac{\gamma(\omega_{D,TT})}{N}(N)P_{T_{n}T_{n+1}}+\sum_{d}\frac{\gamma(\omega_{TT,D})}{N}P_{d}\\
 & +f_{-}\frac{\gamma(\omega_{TT,-})}{N}P_{-}-f_{-}\frac{\gamma(\omega_{-,TT})}{N}P_{T_{n}T_{n+1}}
\end{align*}
where $f_{\pm}^{p}=|\langle1_{ph}|\otimes\langle G|\pm\rangle|^{2}$
and $k_{phot}$ is the rate constant that accounts for photon leakage.
The latter is only relevant for the polariton states as the photonic
component for the dark states is zero. By introducing $P_{D}^{total}=\sum_{d}P_{d}$
and $P_{TT}^{total}=\sum_{n}P_{T_{n}T_{n+1}}$, we arrive at Eqs.
(8a)-(8c) of the main text. We have also introduced the rate constant
$k_{c}(\omega)$ to account for the decay of the singlet due to a
process that competes with SF (see main text). 

\begin{flushleft}
$[R1]$ F. Herrera and F. C. Spano, Phys. Rev. Lett. \textbf{116}, 238301 (2016).\\
$[R2]$ F. C. Spano and H. Yamagata. J. Phys. Chem. B. \textbf{115}, 5133 (2011).\\
$[R3]$ V. May and O. K\"uhn, \textit{Charge and energy transfer
dynamics in molecular systems} (John Wiley \& Sons,
Berlin, 2008).\\
$[R4]$ J. Jortner, J. Chem. Phys. \textbf{64}, 4860 (1976).\\
$[R5]$ N. R. Monahan, D. Sun, H. Tamura, K. W. Williams, B. Xu, Y. Zhong, B. Kumar, C. Nuckolls, A. R. Harutyunyan, G. Chen, H.-L. Dai, D. Beljonne, Y. Rao, and X. Y. Zhu. Nat. Chem. \textbf{9}, 341 (2017).\\
$[R6]$ S. R. Yost, J. Lee, M. W. Wilson, T. Wu, D. P. McMahon, R. R. Parkhurst, N. J. Thompson, D. N. Congreve, A. Rao, K. Johnson, M. Y. Sfeir, M. G. Bawendi, T. M. Swager, R. H. Friend, M. A. Baldo, and T. Van Voorhis, Nat. Chem. \textbf{6}, 492 (2014).\\
$[R7]$ J. Jortner and M. Bixon, J. Chem. Phys. \textbf{88}, 167 (1988).\\
$[R8]$ J. del Pino, J. Feist, and F. J. Garcia-Vidal, New J. Phys. \textbf{17}, 053040 (2015).\\
$[R9]$  T. C. Berkelbach, M. S. Hybertsen, and D. R. Reichman, J. Chem. Phys. \textbf{138}, 114103 (2013).\\
\end{flushleft}

\end{document}